\documentclass[12pt]{iopart}
\usepackage[pdftex]{graphicx}
\usepackage{dcolumn}
\usepackage{bm}
\usepackage{epsfig}
\usepackage{dsfont}
\usepackage{bbm}
\usepackage{amsfonts}
\usepackage{amssymb}
\usepackage{amscd}
\usepackage{amstext}
\usepackage{float}
\usepackage{rotating,standalone}
\usepackage{color}
\usepackage{ulem}

\usepackage{iopams}
\usepackage{setstack}

\usepackage{tikz}

\newcommand{\bra}[1]{\langle\,{#1}\, |}
\newcommand{\ket}[1]{|\,{#1}\,\rangle}

\newcommand{\sub}[2]{{#1}_{ \mbox{\scriptsize #2}}}
\newcommand{\supp}[2]{{#1}^{ \mbox{\scriptsize #2}}}

\def\beq{\begin{eqnarray}}
	\def\eeq{\end{eqnarray}}

\newcommand{\rref}[1]{Ref.~\cite{#1}}

\newcommand{\frefp}[2]{figure~\ref{#1}(#2)}

\newcommand{\cref}[1]{chapter~\ref{#1}}
\newcommand{\Cref}[1]{Chapter~\ref{#1}}
\newcommand{\aref}[1]{\ref{#1}}
\newcommand{\bref}[1]{(\ref{#1})}

\renewcommand{\emph}[1]{\textit{#1}}

\begin{document}
\title{Excitons guided by polaritons}
\author{K.~Mukherjee$^1$ and S.~W\"uster$^1$}
\address{$^1$Department of Physics, Indian Institute of Science Education and Research, Bhopal, Madhya Pradesh 462 066, India}
\ead{sebastian@iiserb.ac.in}
\begin{abstract}
We show that an exciton on a discrete chain of sites can be guided by effective measurements induced by an ambient, non-equilibrium medium that is synchronised to the exciton transport.
For experimental verification, we propose a hybrid cold atom platform, carrying the exciton as electronic excitation on a chain of atoms, which are surrounded by a slow light medium supporting polaritons.
The chain is coupled to the medium through long-range Rydberg interactions. Despite the guiding mechanism being incoherent, the exciton pulse can be coherently transported with high fidelity. The implementation requires careful alignment of chain and medium but then no further time-dependent control. Our concept can be ported to other
exciton and polariton carrying media or devices, and will enable switches and waveguides operating with the two quasi particles involved, as we demonstrate.
\end{abstract}

\maketitle

\section{Introduction}
%
 While computational devices of the 20th century mainly utilized electrons and holes as operational entities, quantum devices of the present century will likely exploit a significantly wider pool, functionalising spinons, polaritons \cite{angelakis2017quantum,ghosh2020quantum}, plasmons \cite{alonso2019quantum} and excitons \cite{de2002intrinsic,chen2001control}, among many other quasi-particles. Future technologies will also leverage quantum coherence, essential for quantum information and quantum computing. However, as we progress towards larger quantum devices, decoherence will become significant \cite{book:schlosshauer,Schlosshauer_decoherence_review}, as will transport of quantum information from one part of the device to another. Although decoherence is typically considered detrimental \cite{Suter_RevModPhys}, it can also be turned into a resource \cite{Poyatos_dissip_engineering_PhysRevLett,Verstraete_dissip_engineering,Vuglar_nonconsforces_PhysRevLett}.
 
Along that line, we develop a design principle for hybrid platforms in which a polariton in a bulk medium coherently guides an exciton on a discrete chain or waveguide embedded in that medium. For this, chain and medium are linked by localized interaction sites, which affect the exciton only when they are reached by the polariton and then guide its transport.
We show that guiding enables high fidelity exciton transport and provides a control interface linking the polariton carrying component and exciton carrying component. This hybrid quantum technology functionalises both types of quasi particles and turns decoherence by the interactions between the components into an asset. 

We find highly coherent and controlled exciton transport, despite the underlying guiding mechanism relying on an effective measurement and hence decoherence. Our concept can be leveraged for a variety of quantum technologies, since the guided transport remains coherent and could thus carry quantum information, while the guiding mechanism is decoherence, for which sources are broadly available.
For that reason, we expect that the present scheme can find applications for example in future electromagnetically induced transparency (EIT) in semi-conductor optics \cite{phillips2003electromagnetically}, Rydberg-excitons in semiconductors \cite{walther2018giant,walther2018interactions},  artificial photosynthetic devices \cite{saikin2013photonics,zhang2017dye,ghosh1978merocyanine,pant2020excitation}, quantum computing \cite{saffman2016quantum}, quantum batteries \cite{yao2021stable}, quantum switches \cite{amo2010exciton} and other solid state based devices.

As a specific example here, that also can be demonstrated with state of the art experiments, we consider a regular chain of Rydberg atoms, a Rydberg aggregate \cite{wuester:review}, embedded in a cold gas medium, while the two are interfaced through suitably distributed detector atoms. These detector atoms can control the exciton transport through measurement induced guiding (MIG), when they are triggered by a suitably designed polariton pulse.

\begin{figure}[htb!]
	\centering
	\includegraphics[width=0.59\linewidth]{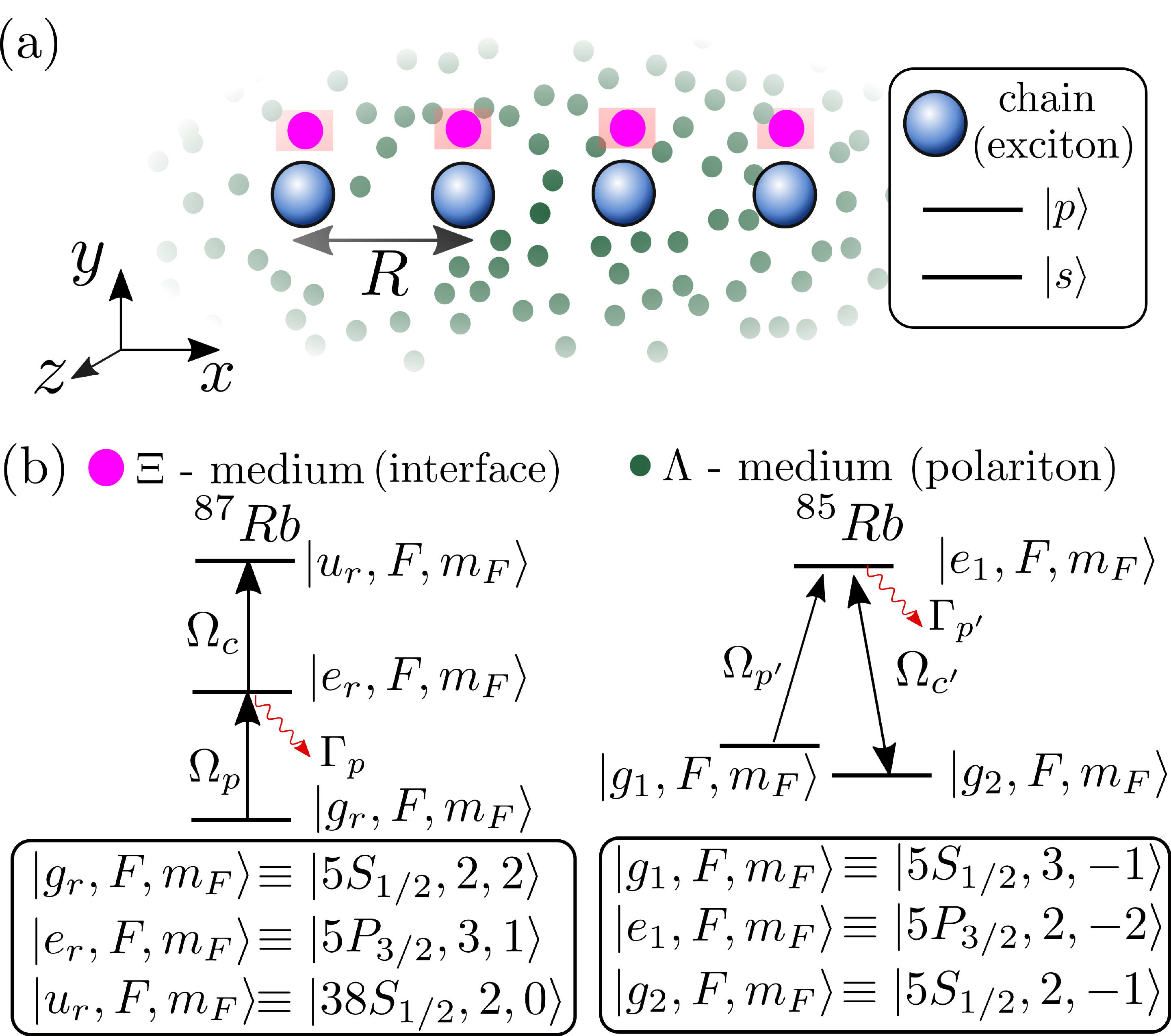}
	\caption{
	Setup for guiding excitons with polaritons. (a) An equidistant chain of Rydberg atoms (large blue-balls), that can be in two internal states as shown on the right, is embedded in background gas
	of carrier atoms (green-balls), that also contains detector atoms as interface (magenta-balls within square). The latter are realized with $^{87}$Rb and the former with $^{85}$Rb, to allow separate control and trapping of the components. (b) In each 
	isotope, we require an effective three level system to realize EIT, shown separately for detector atoms ($\Xi$ - medium) and carrier atoms ($\Lambda$ - medium). Red boxes near detector atoms in (a) represent focussed coupling beams, with Rabi frequency $\Omega_{c}$. 
}
	\label{fig1}
\end{figure}

 The article is structured as follows: we first establish the exemplary physical system in section \ref{sec:system}, discussing the polariton and exciton carrying media. In section \ref{sec:results}, we demonstrate high-fidelity exciton transport guided by polaritons, and in section \ref{subsec:quantum_switch} incoherent switching of exciton transport, before we conclude in section \ref{sec:conclusion}. Detailed level schemes for the proposed implementation, and options for further fidelity enhancement are provided in appendices.

\section{The system}
\label{sec:system}

We consider an atomic gas composed of the two Rubidium isotopes, $^{85}$Rb and $^{87}$Rb. Each is used for a different purpose, with $^{87}$Rb contributing two types of atoms: (i) \textit{aggregate} atoms, shown as blue balls in \fref{fig1}~(a), that will carry the exciton and (ii) \textit{detector} atoms (magenta balls). Both are trapped at specified locations, providing the interface between excitons and polaritons. The third type of atoms, \textit{carrier} atoms (green balls), are distributed as an untrapped bulk $^{85}$Rb gas surrounding the first two types of trapped atoms, and will carry the polaritons.

The $N$ aggregate atoms are excited into a Rydberg state, forming a Rydberg aggregate \cite{wuester:review}, evenly spaced with separation $R$. Each is flanked by a detector atom as shown. For aggregate atoms, we consider only two Rydberg states: $\ket{s}\equiv\ket{n=43;l=0}$ and $\ket{p}\equiv\ket{n=43;l=1}$, and then further restrict the many-body Hilbert-space to contain just a single $p$-excitation on the chain.
The resulting many-body basis is $\ket{\pi_n}=\ket{ss..p..s}$, where the $n$'th atom is in the Rydberg-$p$ state and the remaining aggregate atoms are in the Rydberg-$s$ state. 

The heart of the proposed platform is the separate implementation of ladder-type EIT \cite{fleischhauer:review,friedler:longrangepulses,Mohapatra:coherent_ryd_det,mauger:strontspec,Mohapatra:giantelectroopt,schempp:cpt,sevincli:quantuminterf,parigi:interactionnonlin,srakaew2022subwavelength} ($\Xi$-medium) for the detector atoms and lambda-type EIT \cite{fleischhauer:review} ($\Lambda$-medium) for the carrier atoms, as illustrated in \fref{fig1}~(b). To describe these, we use the electronic basis $\{g_r,e_r,u_r\}$ for the detector atoms (EIT-$\Xi$) and $\{g_1,e_1,g_2\}$ for the carrier atoms (EIT-$\Lambda$). 
We require $\ket{u_r}$ to be a strongly interacting state that will couple excitons and polaritons. In our example, where we consider two atomic species forming the $\Xi$ and $\Lambda$ media, respectively, this is realized when the upper level of EIT-$\Xi$ is a Rydberg state. Specific hyperfine states proposed for those respective atomic species are indicated in \fref{fig1}~(b). The exemplary states are selected such that $\ket{g_r}\rightarrow\ket{e_r}$ and $\ket{g_1}\rightarrow\ket{e_1}$ can be coupled using the same optical field, which is the case here when both transition frequencies are made equal with an additional Zeeman shift as discussed in \aref{app:level_scheme}. Subsequent transitions in the two level schemes however have to be species selective.

The specific proposal of an atomic system above is for concreteness and direct experimental demonstration, while the design principle introduced is more general. For example EIT is generic to three level systems, occurring also in optomechanics \cite{Painter_Nature_EIT_optomech} or semi-conductor quantum wells \cite{Phillips_EIT_quantwells_PRL}, while the chain could be a quantum dot array \cite{kagan1996long} instead of atoms. 

The setup hybridizes two components: Firstly the aggregate, which allows a (Frenkel) \textit{exciton} in the form of the Rydberg $p$-excitation to be transported coherently in absence of environmental interactions \cite{wuester:review,barredo2015coherent}. This excitation transport can be significantly affected by controlled decoherence through an effective measurement made by the detector atoms \cite{David:Rydagg,schempp:spintransport} as discussed in \ref{subsec:MID}. Secondly, the background gas, which can carry \textit{polaritons} formed through the light-matter interaction within EIT-$\Lambda$ involving the carrier atoms. This polariton transport is tunable using EIT-$\Lambda$ parameters, as reviewed and applied in \sref{subsec:EITslowlight}.

 \subsection{Exciton dynamics}
\label{subsec:MID}
To model the exciton carrying component, the Hamiltonian of a Rydberg aggregate describing dipole-dipole interactions is given by 
	\begin{eqnarray}
		\sub{\hat{H}}{agg} = \sum_{n,m\neq n}^N\frac{C_3}{|\mathbf{r}_n-\mathbf{r}_m|^3}\ket{\pi_n}\bra{\pi_m},
		\label{Hagg}
	\end{eqnarray}
	where $C_3$ is the dipole-dipole dispersion coefficient and $\mathbf{r}_n$ the position of the $n$th Rydberg atom in the aggregate. The Hamiltonians to describe EIT of the detector and carrier atoms respectively, after the dipole and rotating wave approximation, are
	\begin{eqnarray}
		\hat{H}_{\Xi}& =& \sum_\alpha^{\sub{N}{det}} \frac{\Omega_p(\mathbf{x}_\alpha,t)}{2}\hat{\sigma}^{(\alpha)}_{g_r e_r}+\frac{\Omega_{c}}{2}\hat{\sigma}^{(\alpha)}_{e_r u_r}+h.c. ,\label{EIT-Xi}\\
		\hat{H}_{\Lambda}& =& \sum_\beta^{\sub{N}{car}}  \frac{\Omega_{p'}(\mathbf{x}_\beta,t)}{2}\hat{\sigma}^{(\beta)}_{g_1e_1}+\frac{\Omega_{c'}(\mathbf{x})}{2}\hat{\sigma}^{(\beta)}_{e_1g_2}+h.c., \label{EIT-Lambda}
	\end{eqnarray}
	where $\hat{\sigma}^{(\zeta)}_{kl}=[\ket{k}\bra{l}]_\zeta$ for $\zeta\in\{\alpha,\beta\}$ acts on the $\alpha$th \textit{detector} atom (or $\beta$th \textit{carrier} atom) only, the former is at position $\mathbf{x}_\alpha$, and $\Omega_p$ ($\Omega_{c}$) are the probe (coupling) Rabi frequency for detector atoms (ladder EIT, $\Xi$-medium), while $\Omega_{p'}$ ($\Omega_{c'}$) are those for carrier atoms (Lambda EIT, $\Lambda$-medium). $\Omega_p$ and $\Omega_{p'}$ should involve the same transition frequency, such that the same field can drive $\ket{g_r}\rightarrow\ket{e_r}$ and $\ket{g_1}\rightarrow\ket{e_1}$ simultaneously. This will enable a polariton propagating in the $\Lambda$-medium to trigger a detector atom.
The coupling beam $\Omega_{c}$ is tightly focused onto the detector atoms only, in order to prevent any conflict between the two EIT schemes, as indicated by pink shades in \frefp{fig1}{b}. In addition, our exemplary choice of states also makes sure that the transitions $\ket{e_1}\rightarrow\ket{g_2}$ and $\ket{e_r}\rightarrow\ket{u_r}$ are only driven for the respective target species, see \aref{app:level_scheme}.
		
Spontaneous decay of states involved in EIT is modelled using decay operators $\hat{L}^\zeta_{kl}=\sqrt{\Gamma_{kl}}\hat{\sigma}_{kl}^{(\zeta)}$ with $\zeta\in\{\alpha,\beta\}$ and $\Gamma_{kl}$ the spontaneous decay rate of level $\ket{l}$ to $\ket{k}$. Here, we only explicitly require the decay operator relevant for EIT-$\Xi$, which is $\hat{L}_\alpha = \sqrt{\Gamma_p} \hat{\sigma}_{g_re_r}^{(\alpha)}$ with $\Gamma_p\equiv\Gamma_{g_re_r}$.
 
 Since the upper level of EIT-$\Xi$ is a Rydberg state $\ket{u_r}\equiv \ket{38;l=0}$, the detector atoms realize Rydberg-EIT 
\cite{fleischhauer:review,friedler:longrangepulses,Mohapatra:coherent_ryd_det,mauger:strontspec,Mohapatra:giantelectroopt,schempp:cpt,sevincli:quantuminterf,parigi:interactionnonlin}
and experience long-range van der Waals interactions with Rydberg aggregate atoms. 
The interaction Hamiltonian $\sub{H}{int}$ accounting for these is
	\begin{eqnarray}
		\sub{\hat{H}}{int} = \sum_{n,\alpha}\bar{V}_{\alpha n}[\ket{u_r}\bra{u_r}]_\alpha\ket{\pi_n}\bra{\pi_n},
		\label{H_int}
	\end{eqnarray}
	where $\bar{V}_{\alpha n}=V^{\alpha n}_{u_rp}+\sum_{m\neq n}V^{\alpha m}_{u_rs}$ based on $V^{\alpha n}_{u_ra} =C_{\eta(a),u_ra}/|\mathbf{r}_\alpha - \mathbf{r}_n|^{\eta(a)} $ with $\eta(a)=6,4$ for  $a=s,p$, respectively, is the net interaction of the $\ket{u_r}$ state with the aggregate in $\ket{\pi_n}$ \cite{David:Rydagg}. Here, $\mathbf{r}_\alpha$ and $\mathbf{r}_n$ are the position of the $\alpha$th detector atom and the $n$th Rydberg atom. Since the probability of two detector atoms being simultaneously excited to $\ket{u_r}$ will remain small, interactions between detector atoms are negligible \cite{David:Rydagg,olmos:amplification}. Carrier atoms (EIT-$\Lambda$) are only present in low lying states, hence interactions between Rydberg atoms and carrier atoms (EIT-$\Lambda$) can be ignored in comparison to those between Rydberg states.
	
	We describe the dynamics of the system incorporating decoherence effects from EIT-$\Xi$ with the Lindblad master equation ($\hbar=1$)
	\begin{eqnarray}
		\dot{\hat{\rho}}=-i[\hat{H},\hat{\rho}]+\sum_\alpha \mathcal{L}_{\hat{L}_\alpha}[\hat{\rho}], 
		\label{LB_full}
	\end{eqnarray}
	where $\hat{H}$ is the total Hamiltonian, $\hat{\rho}$ is the density operator and $\mathcal{L}_{\hat{L}_\alpha}$ is a superoperator, defined in the form $\mathcal{L}_{\hat{O}}=\hat{O}\hat{\rho}\hat{O}^\dagger-(\hat{O}^\dagger\hat{O}\hat{\rho}+\hat{\rho}\hat{O}^\dagger\hat{O})/2$. 
	
	To reduce the dimension of the Hilbert-space and reach a simpler master equation, we use an effective model described in \cite{David:Rydagg}, derived using the effective operator formalism of \rref{Sorenson:eff_operator}. It allows a description of dynamics in the space of the aggregate alone, using a reduced density matrix $\supp{\hat{\rho}}{agg}=\sum_{nm}\rho_{nm}\ket{\pi_n}\bra{\pi_m}$, evolving according to 
	\begin{eqnarray}
		\supp{\dot{\hat{\rho}}}{agg}&=-i[\sub{\hat{H}}{agg} + \sub{\hat{H}}{eff},\supp{\hat{\rho}}{agg}]+\sum_\alpha \mathcal{L}_{\sub{\hat{L}}{eff}^{(\alpha)}}[\supp{\hat{\rho}}{agg}], 
		\label{LB_eff}
	\end{eqnarray}
	where  $\sub{\hat{H}}{agg}$ is defined in \eref{Hagg}, and the effective Hamiltonian and the Lindblad operator incorporating $\sub{\hat{H}}{int}$, $\hat{H}_{\Xi}$ and $\hat{L}_\alpha$ \cite{David:Rydagg} are written as 
	\begin{eqnarray}
		\sub{\hat{H}}{eff} &=&\sum_n\bigg[\sum_\alpha \frac{\Omega_p^2(\mathbf{x}_\alpha,t)}{\Omega_c^2}\frac{\bar{V}_{\alpha n}}{1+(\bar{V}_{\alpha n}/V_c)^2}\bigg]\ket{\pi_n}\bra{\pi_n} ,\label{Heff}\nonumber\\
		\sub{\hat{L}}{eff}^{(\alpha)} &=&-\sum_n\bigg[\frac{\Omega_p(\mathbf{x}_\alpha,t)}{\sqrt{\Gamma_p}} \frac{1}{i+V_c/\bar{V}_{\alpha n}}\bigg]\ket{\pi_n}\bra{\pi_n},
		\label{Leff}
	\end{eqnarray}
	where $V_c=\Omega_{c}^2/(2\Gamma_p)$. The operators $\sub{\hat{L}}{eff}^{(\alpha)}$ describe decoherence that depends on the aggregate state $\ket{\pi_n}$ and atomic positions, 
	caused by the fact that van der Waals interactions can break EIT-$\Xi$ by shifting the state $\ket{u_r}$ out of resonance, as first proposed in \cite{guenter:EITexpt} and demonstrated in \cite{Gunter:EIT}.
	
Away from the aggregate Rydberg atoms, EIT-$\Xi$ turns the background gas transparent for the probe beam. The presence of embedded Rydberg atoms, due to the interaction energy shift discussed above, then give rise to two main features: (i) van der Waals interactions between Rydberg atoms and detector atoms disrupt EIT-$\Xi$ at close proximity, generating an absorption shadow 
of radius $R_{c,s}$ around a Rydberg atom in state $\ket{s}$ and of radius $R_{c,p}$ around a Rydberg atom in state $\ket{p}$.
The radii are $R_{c,a}=(2C_{\eta(a),ra}\Gamma_p/\Omega_{c}^2)^{1/\eta(a)}$ where $\eta(a)=6,4$ for $a=s,p$, and can be tuned to significantly differ \cite{David:Rydagg}. The shadows indicate the presence of Rydberg atoms and indirectly measure their location (and state) via the centre (and size) of the shadow, implementing a quantum non-demolition (QND) measurement of both properties. (ii) Controllable decoherence can thus be induced in the excitation transport by positioning detector atoms between the shadow radii $R_{c,s}$ and $R_{c,p}$. Then they can measure the location of the Rydberg-$p$ excitation on the aggregate, and thus the state $\ket{\pi_n}$. The resultant effect on the quantum state of the aggregate can give rise to tunable measurement-induced guiding (MIG), presented in \sref{sec:results}.

	\subsection{Tuned polariton propagation}
	\label{subsec:EITslowlight}
\begin{figure}
	\centering
	\includegraphics[width=0.59\linewidth]{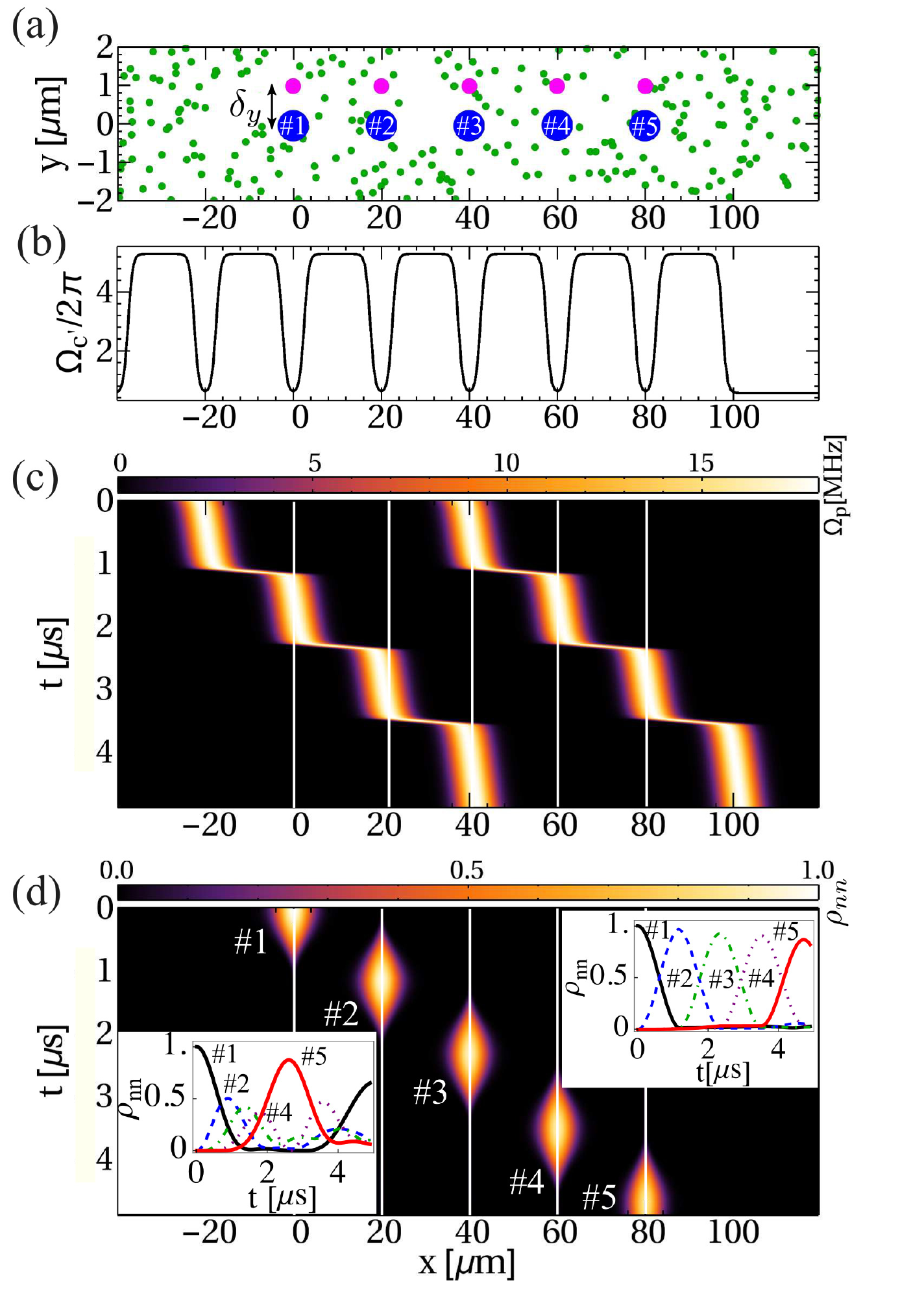}
	\caption{(a) Array of $5$ Rydberg aggregate atoms (blue) spaced by $R=20\mu$m, flanked by detector atoms (magenta), all trapped in individual harmonic traps with ground-state width $\sub{\sigma}{trap}=0.5\mu$m. All are immersed in a background gas of $^{85}$Rb carrier atoms at uniform density $\varrho_{85} = 3.75\times 10^{13}$m$^{-2}$. 
		(b) The coupling field profile $\tilde{\Omega}_{c'}(x)$ as in \eref{Omegacp}, using a rise-length $\sigma_c=1.0\mu$m and $\Omega_{c',min}/2\pi= 0.5$ kHz, $\Omega_{c',max}/2\pi=5.28$ kHz.  (c) Resultant propagation of two probe pulses in $\Omega_p(\mathbf{x},t)$ with Gaussian envelope of width $\sigma_p = 3\mu m$.  White vertical lines in (c-d) indicate the position of aggregate atoms marked with the atom index $n$.
		(d) Modelled using \eref{LB_eff}, an excitation initialised at site 1, at $x=0$ $\mu$m, is transported to site 5, at $x=80$ $\mu$m, via site 2,3 and 4 with high fidelity, enabled by synchronized activation of detector atoms and polariton transport shown in (c). We indicate the excitation population $\rho_{nn}$ of each atom by the width and amplitude of the colour shading surrounding the white vertical lines. The right inset in (d) again shows the atomic populations $\rho_{nn}$ for the guided case, and the left inset for the isolated aggregate that has no surrounding medium. Parameters used are $C_3/2\pi=1619$ MHz $\mu$m$^3$, $C_{6,s}/2\pi=-87$ MHz $\mu$m$^6$, $C_{4,p}/2\pi=-1032$ MHz $\mu$m$^4$, $\Omega_{p,0}/2\pi=2.5\Omega_{p',0}/2\pi=18$ MHz, $\Omega_{c}/2\pi=90$ MHz and $\Gamma_{p}/2\pi=6.1$ MHz.
		\label{transport}}
\end{figure}

In the second component of the hybrid platform, we now discuss how EIT-$\Lambda$ on the carrier atoms can control the motion of the probe field polaritons \cite{fleischhauer2000dark}, to synchronize MIG with dipolar interactions between the Rydberg atoms. For this purpose, writing $\mathbf{x}=[x,y,z]^T$, we consider a spatially dependent coupling field for EIT-$\Lambda$ 
	\begin{eqnarray}
		\Omega_{c'}(\mathbf{x}) &=& \Omega_{c',min}+ \frac{(\Omega_{c',max}-\Omega_{c',min})}{2}  \sum_{n=-2}^{N+1} \bigg(\tanh\left[\frac{x -(n - 1/2)R +w}{\sigma_c}\right] \nonumber\\
		& &- \tanh\left[\frac{x - (n - 1/2)R - w}{\sigma_c}\right] \bigg),
		\label{Omegacp}
	\end{eqnarray}
	 shown in \fref{transport}~(b). In \eref{Omegacp}, $\Omega_{c',min}$ and $\Omega_{c',max}$  are the minimum and maximum Rabi frequency, $\sigma_c$ controls the spatial sharpness of the increasing and decreasing flanks of the profile, $N$ is the number of aggregate atoms and $R$ is the separation between them. We have taken the half width of square waveforms as $w=3R/8$. The coupling field $\Omega_{c'}(\mathbf{x})$ varies along the $x$-component of $\mathbf{x}$ only, and for subsequent equations in one dimension we shall thus use the notation $\Omega_{c'}(\mathbf{x})\rightarrow\tilde{\Omega}_{c'}(x)$, for simplicity. Complex beam shapes such as \bref{Omegacp} can be created using e.g.~spatial light modulators \cite{efron1994spatial}.
The polaritons formed in EIT-$\Lambda$, can be written as \cite{fleischhauer2000dark},
	\begin{eqnarray}
		\hat{\Psi}(x,t)&=&\cos \theta(x)\hat{E}(x,t)- \sin\theta(x)\sqrt{\varrho_{85}}\hat{\sigma}_{gg_1}(x,t),\\
		\cos\theta(x)&=& \frac{\tilde{\Omega}_{c'}(x)}{\sqrt{\tilde{\Omega}_{c'}^2(x) + g^2\varrho_{85}}}. \label{cos_theta} 
	\end{eqnarray}
Here, $\hat{E}$ is a quantum field resonantly coupling the states $\ket{g_1}\leftrightarrow\ket{e_1}$ with coupling constant $g=\mu \sqrt{\omega_0/2\pi \epsilon_0}$, where $\mu$ and $\omega_0$ are the transition dipole matrix element and transition frequency, respectively. $\varrho_{85}$ is the background gas density of $^{85}$Rb carrier atoms.
As is well known, $\hat{\Psi}$ obeys the equation of motion 
	\begin{eqnarray}
		\bigg[\frac{\partial}{\partial t} + c \cos^2[\theta(x)]\frac{\partial}{\partial x} \bigg] \hat{\Psi}(x,t)=0,
		\label{eom_polariton}
	\end{eqnarray}
	which describes a shape-preserving wave propagation with group velocity $v_g(x)=c \cos^2[\theta(x)]$, where $c$ is the vacuum speed of light. Since $\cos \theta(x)$ decreases with $\tilde{\Omega}_{c'}(x)$ in \eref{cos_theta}, the group velocity of the polariton decreases with the strength of $\tilde{\Omega}_{c'}(x)$, giving rise to slow light \cite{fleischhauer:review}. Note, that the  functionality of the detector atoms in EIT-$\Xi$ requires $\Omega_c\gg\Omega_p$, while slowing down polaritons in EIT-$\Lambda$ requires small $\Omega_{c'}$, thus necessitating two different atomic species for EIT-$\Xi$ and EIT-$\Lambda$.
	
	 In the classical limit, the field part of the propagating polariton gives rise to a probe Rabi frequency $\Omega_{p'}(\mathbf{x},t)=g\sqrt{\varrho_{85}}\langle\hat{E}(x,t)\rangle $, which we can use in \eref{EIT-Lambda}. Importantly, since we have designed the probe transition frequencies for EIT-$\Lambda$ and EIT-$\Xi$ to be equal, these polaritons will trigger the detector atoms and we take 
$\Omega_{p}(\mathbf{x},t)\approx2.5\Omega_{p'}(\mathbf{x},t)$, where the proportionality factor arises due to different transition dipole matrix elements. 
	
Incorporating the mean field polariton propagation dictated by \eref{eom_polariton}, we thus assume for $\Omega_p(\mathbf{x},t)$ a train of Gaussian pulses that propagate with group-velocities locally controlled by $\tilde{\Omega}_{c'}(x)$ according to
\begin{eqnarray}
	\Omega_{p}(\mathbf{x},t) = \sum_n \Omega_{p,0}\exp\bigg(-\frac{(x-x_n(t))^2}{2\sigma_p^2}\bigg),
	\label{omegap}
\end{eqnarray}
with $\mathbf{x}=[x,y,z]$, 
$x_n(t)=x_n(0)+ \int_0^t v_g(x_n(t'))dt'$  the instantaneous x-position of the $n$th probe pulse (polariton),
$\Omega_{p,0}$ its peak Rabi frequency and $\sigma_p$ its spatial width. The corresponding profile $\Omega_{p}(\mathbf{x},t)$ along the $x$-axis is shown in \frefp{transport}{c}. Since the profile $\tilde{\Omega}_{c'}(x)$ in \eref{Omegacp} shall later vary on scales $\sigma_c\approx\sigma_p/3$
the pulses will in reality be distorted during their fast transits from detector atom to detector atom, but should remain Gaussian shaped while remaining at one detector atom, which is where they matter for our simulations.

\section{Polaritons guide excitons}
\label{sec:results}

To demonstrate how the combination of effective measurement through interactions and polariton transport discussed in the previous sections gives rise to 
measurement induced guiding (MIG) of the exciton, we consider a Rydberg aggregate of $N=5$ atoms, equidistantly separated by $R=20\mu$m and immersed in a background gas, as shown in \fref{transport}~(a).  The life-time of the aggregate is $17.8$ $\mu$s \cite{beterov:BBR}. Each Rydberg atom is accompanied by a detector atom, offset by $\delta_y=1$ $\mu$m in the y-direction only. The carrier atoms (green balls) are uniformly distributed in the $x-y$ plane shown in \fref{transport}~(a).
	 
We assume the time-independent spatial coupling profile $\tilde{\Omega}_{c'}(x)$ in \eref{Omegacp}, shown in \fref{transport}~(b), and insert two strongly occupied polariton pulses at $x_1=-21.5$ $\mu$m and  $x_2=38.5$ $\mu$m into the probe field, as shown in \fref{transport}~(c). The precise location of polaritons is important to ensure synchronization with exciton transport and thus a high fidelity of the latter. The aggregate is initialised with exciton localised on the first site: $\hat{\rho}=\ket{\pi_1}\bra{\pi_1}$, and the system is then evolved following the effective master equation in \eref{LB_eff}, assuming the propagating profile for the polaritons given in \eref{omegap}, with results shown in \fref{transport}~(d).
	 
We see in panel (d) and its top right inset, that the exciton is guided to the final site with high fidelity, kept localized on at most two sites, in contrast to the free propagation on the isolated aggregate chain, shown in the bottom left inset. 	 
To understand this, let us consider the process step by step: In the first stage, from $t=0$ to $t\approx1.23$ $\mu$s ($\approx \pi R^3/2C_3)$, the polaritons are slowed down by the weak coupling amplitude $\tilde{\Omega}_{c'}$ 
near their initial positions at $x_1\approx-20\mu$m and $x_2\approx40\mu$m. During this period they only activate the detector atom flanking the third site. This results in continuous strong measurement of the third aggregate atom through the nearby detector atom. That atom thus strongly ``measures'' whether the adjacent aggregate atom carries the exciton, freezing its state in the $\ket{s}$ state though a quantum Zeno effect \cite{misra:quantumzeno,Kofman:antizeno}. The energy shift in $\sub{\hat{H}}{eff}$ linked to the detector atom is not significant here, which we verify by setting it to zero, finding nearly unchanged dynamics.
Having thus effectively removed $\ket{\pi_3}$ from the Hilbert space through the Zeno effect, the exciton is transferred by a coherent oscillation to site $\ket{\pi_2}$. This is the essence of measurement induced guiding.
The EIT-$\Lambda$ coupling profile is designed such that exactly after this transfer, the polariton pulses leave their respective slow zone, make a quick transit to the next site within $t\approx2ns$, and now block site 1 and 4, thus nextly allowing a complete exciton transfer from site 2 to site 3, and so forth.

Repeating these steps, we demonstrate a transport fidelity $\rho_{55}$ of up to $87\%$ in \fref{transport}~(d). The simulation did include position uncertainties of aggregate atoms and detector atoms, assuming each to be harmonically trapped with width $\sub{\sigma}{trap}=0.5\mu$m.\footnote{Our proposal assumes a uniform distribution of carrier atoms, although these do not explicitly enter simulations.} We then averaged over $10^3$ realisations of aggregate and detector atom positions. The fidelity can be enhanced further to over $95\%$ with four polaritons, which can also suppress next-to-nearest neighbour dipole interactions between aggregate atoms, see \aref{app:four_polaritons}. Furthermore, the design can be extended to a 2D lattice with $N\times N$ aggregate atoms, where the exciton can be guided using a two-dimensional spatially dependent coupling field $\tilde{\Omega}_{c}(x,y)$.

Note, although the QND measurement that guides transport implies a decohering process,  the transport itself remains coherent with purity $P=\mathrm{Tr}[\sub{\hat{\rho}}{agg}(t)^2]$ dropping only marginally to $P\approx0.8$, 
similar to the drop in fidelity. For the case with four polaritons, the purity remains near $P>0.96$ resulting in effectively coherent transport, despite the process relying on decoherence.

 \section{Incoherent switching of coherent exciton transport}
 \label{subsec:quantum_switch}
 \begin{figure}
 	\centering
 	\includegraphics[width=0.59\linewidth]{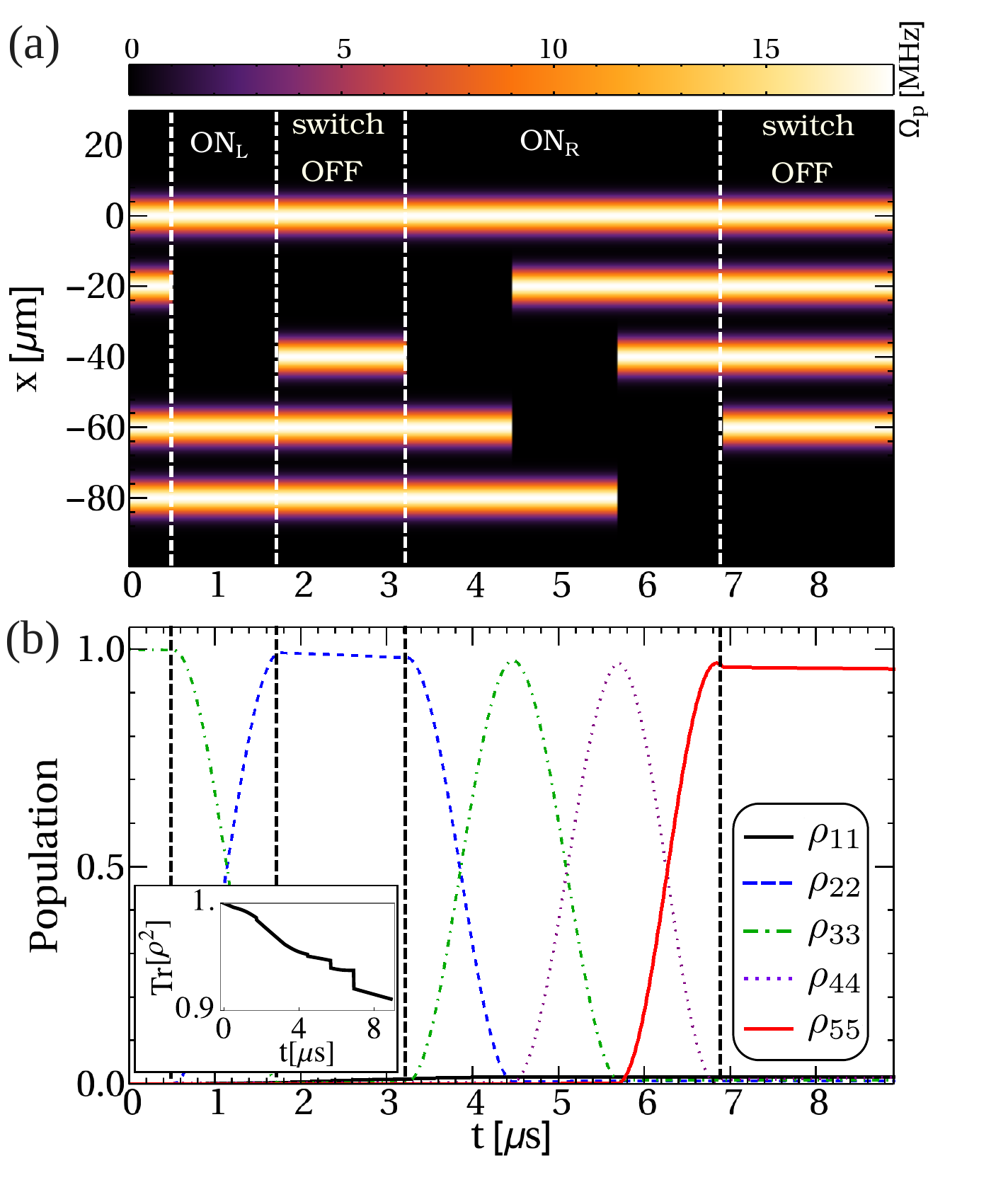}
 	\caption{(a) Probe field $\Omega_p(\mathbf{x},t)$ with set of instructions for controlled excitation switching in the system. (b) Excitation probability $\rho_{nn}$  of aggregate atom $n$. The exciton is initialized on atom $3$ and evolved using \eref{LB_eff}, with the exciton transport controlled by polaritons using switch operations. Other parameters are the same as in \fref{transport}. The vertical lines indicate the timing of implementation of each operation, labelled at the top of (a). The inset in (b) shows the purity $P=\mathrm{Tr}[\sub{\hat{\rho}}{agg}(t)^2]$}
 	\label{switch}
 \end{figure}
 Even without the feature of synchronised propagating polaritons, the MIG through detector atoms can be used to orchestrate complex transport sequences on the Rydberg aggregate.
 As an example, we show in \fref{switch} how it can be used to move the exciton one site to the left, and then two sites to the right.

For this, we consider a set of time-dependent pulses for the probe field $\Omega_p$, shown in \fref{switch}~(a). The probe pulses can be manually switched ``on" and ``off", to carry out a set of instructions that enable or disable exciton transport in a certain direction, thus implementing bi-directional guiding of the exciton within the aggregate. 
  For the demonstration, we initialise the array of $N=5$ Rydberg atoms in $\hat{\rho}=\ket{\pi_3}\bra{\pi_3}$, which would result in its spreading in both directions $-x$ / $+x$ in the unguided aggregate.
Instead we implement three operations: (i) switch OFF, to hold the excitation on a single atom, (ii) switch ON$_L$, only allowing coherent excitation transport by one site to the left (in the $-x$ direction), and (iii) switch ON$_R$, similar to ON$_L$ but transporting to the right (in the $+x$ direction). 
  
 We again see in \fref{switch}, that the purity remains close to $0.95$ after these operations. The switch off feature can have a flexible duration, only limited by the life-time of the system, but the timing of switch OFF and switch ON needs to be precise and depends on the Rabi period during which the excitation transfers coherently from one site to a neighboring site, given by $\sub{t}{period}=\pi R^3/2C_3$. 
   
In more complex scenarios, such as considering two dimensions, the switching can again be induced by propagating and then slowed down polariton pulses as in \sref{sec:results}, providing a hybrid interface between excitonic and polaritonic platforms \cite{ghosh2020quantum}, with possible applications in measurement based quantum computing \cite{briegel2009measurement}.

	\section{Conclusion}
	\label{sec:conclusion}
	We have shown that guided exciton transport is possible in hybrid platforms that carry excitons and polaritons. The guiding is based on continuous Zeno-like measurements of the exciton-carrying component, that are synchronised with transport by propagating polaritons in a guidance medium. We demonstrate the proposed concept explicitly, by modelling a platform compatible with state-of-the-art experiments, in which Rydberg atoms, as exciton carrying component, are embedded in a cold gas medium involving two separate implementations of EIT, the latter providing the polariton-carrying component. 
	
	 It is known that an environment can be beneficial for transport, despite decohering it, for example in exciton migration through molecular aggregates relying on intra-molecular vibrations \cite{renger:review,Foe48_55_,Foe65_93_,AmVaGr00__}, which was recently more broadly discussed 
		\cite{rebentrost2009environment,kassal2012environment,maier2019environment}. In stark contrast to these, our proposal uses an environment out of equilibrium, and then does \emph{not decohere} the transport, while enhancing it.
	
	Beyond the cold atom demonstration here, the design principle proposed will apply to Rydberg-excitons in semiconductors \cite{walther2018giant,walther2018interactions},  artificial photosynthetic devices \cite{saikin2013photonics,zhang2017dye,ghosh1978merocyanine,pant2020excitation}, quantum computing \cite{saffman2016quantum}, quantum batteries \cite{yao2021stable}, quantum switches \cite{amo2010exciton} and other solid state based devices. The key feature to guiding excitons by polaritons that then has to be ported to these systems, is that the guiding medium accesses a strongly interacting state \emph{only} in the presence of the polariton, which then triggers decoherence in the discrete component to guide the exciton.

	\ack
	We acknowledge helpful discussions with Sourav Dutta, Ashok Mohapatra, Swayangdipta Bera, Abhijit Pendse and Aniruddha Mitra. We gladly thank Shannon Whitlock for discussions and reading the manuscript. We also thank the Max-Planck society for financial support under the MPG-IISER partner group program as well as the Indo-French Centre for the Promotion of Advanced Research - CEFIPRA for funding. K.M.~acknowledges the Ministry of Education for the Prime Minister's Research Fellowship (PMRF).
	
\appendix

\section{Level scheme}
\label{app:level_scheme}
	\begin{figure}[htb]
		\centering
	\includegraphics[width=0.89\linewidth]{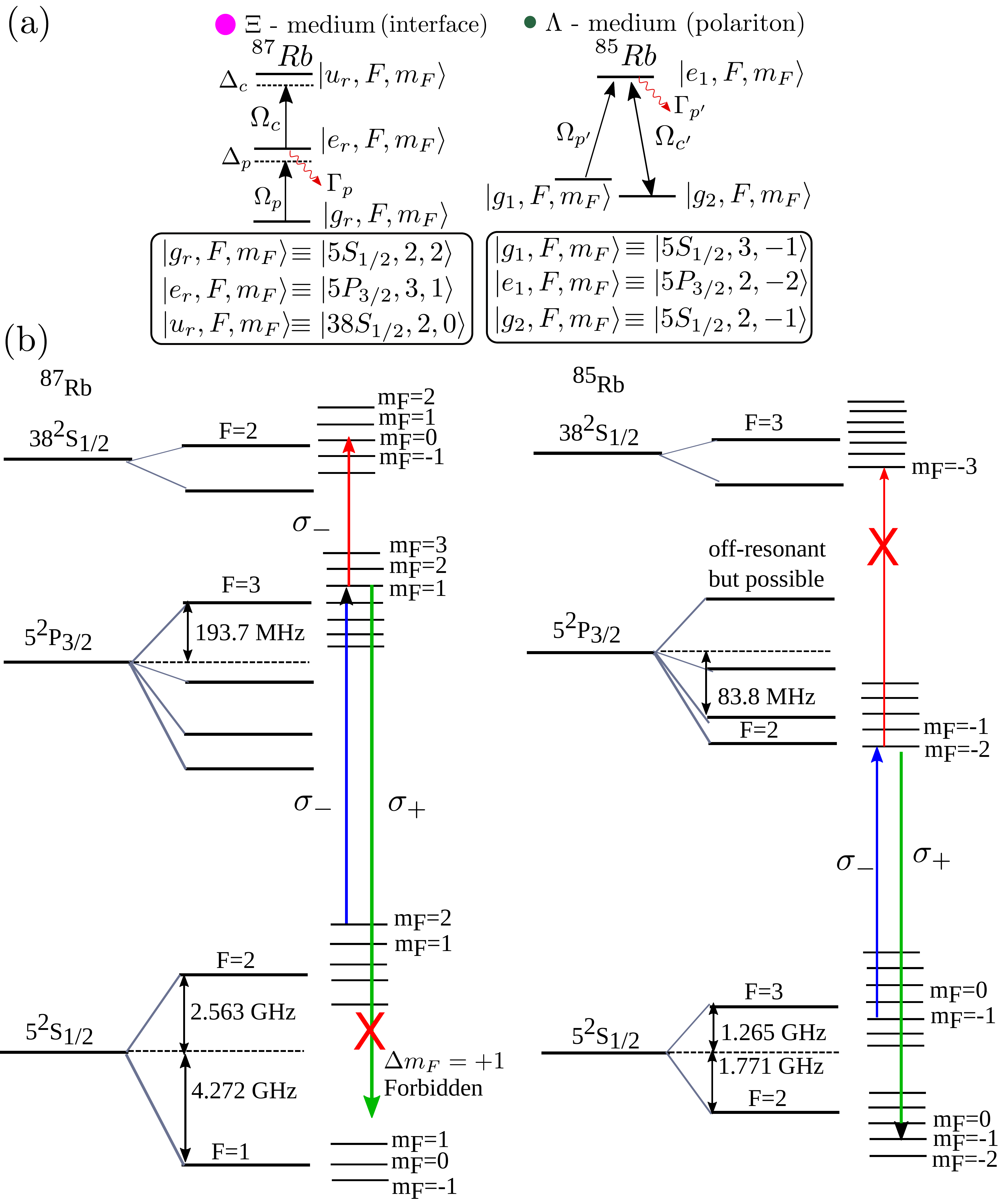}
	\caption{  (a) Energy levels for each EIT scheme as depicted in \fref{fig1}~(c). (b) More detailed energy level diagram for $^{87}$Rb (left) and $^{85}$Rb (right) showing hyperfine levels and their corresponding Zeeman shift for the levels in consideration here \cite{steck2001rubidium85,steck2001rubidium87}. }
	\label{level_scheme}
\end{figure}

A more detailed energy level diagram for $^{87}$Rb \cite{steck2001rubidium87} and $^{85}$Rb \cite{steck2001rubidium85} than the one sketched in \fref{fig1} is shown in \fref{level_scheme}~(b),
 including hyperfine levels and Zeeman shifts. As discussed in the main text, the probe field (blue arrows) drives the transitions $\ket{g_r}\rightarrow\ket{e_r}$ in $^{87}$Rb and $\ket{g_1}\rightarrow\ket{e_1}$ in $^{85}$Rb. In the absence of a magnetic field, these transition are detuned by $\Delta_p=\epsilon_{e_rg_r}-\epsilon_{e_1g_1}\approx1.3$ GHz. However, the difference in the transition energies ($\Delta_p$) can be tuned to zero by Zeeman shifting the levels using a magnetic field of $B_z = 20.12$ mT, making $\Delta_p=0$. This is a strong magnetic field, but within range of typical Feshbach coils used in similar experiments. In addition, we also make sure the coupling field of EIT-$\Xi$ (red arrow) does not interact with EIT-$\Lambda$ by making sure there is no state at the energy of $\ket{u_r}$ in EIT-$\Lambda$ and only focussing that coupling beam onto detector atoms. Also the coupling field of EIT-$\Lambda$ (green arrow) does not interfere with EIT-$\Xi$ due to the non-existence of a $|5S_{1/2}F=1,m_F=2\rangle$ state, thereby making the $\sigma_+$ transition forbidden for any other state of $^{87}$Rb.

\section{Enhancing fidelity using four polaritons}
\label{app:four_polaritons}
%
 \begin{figure}[htb]
	\centering
	\includegraphics[width=0.69\linewidth]{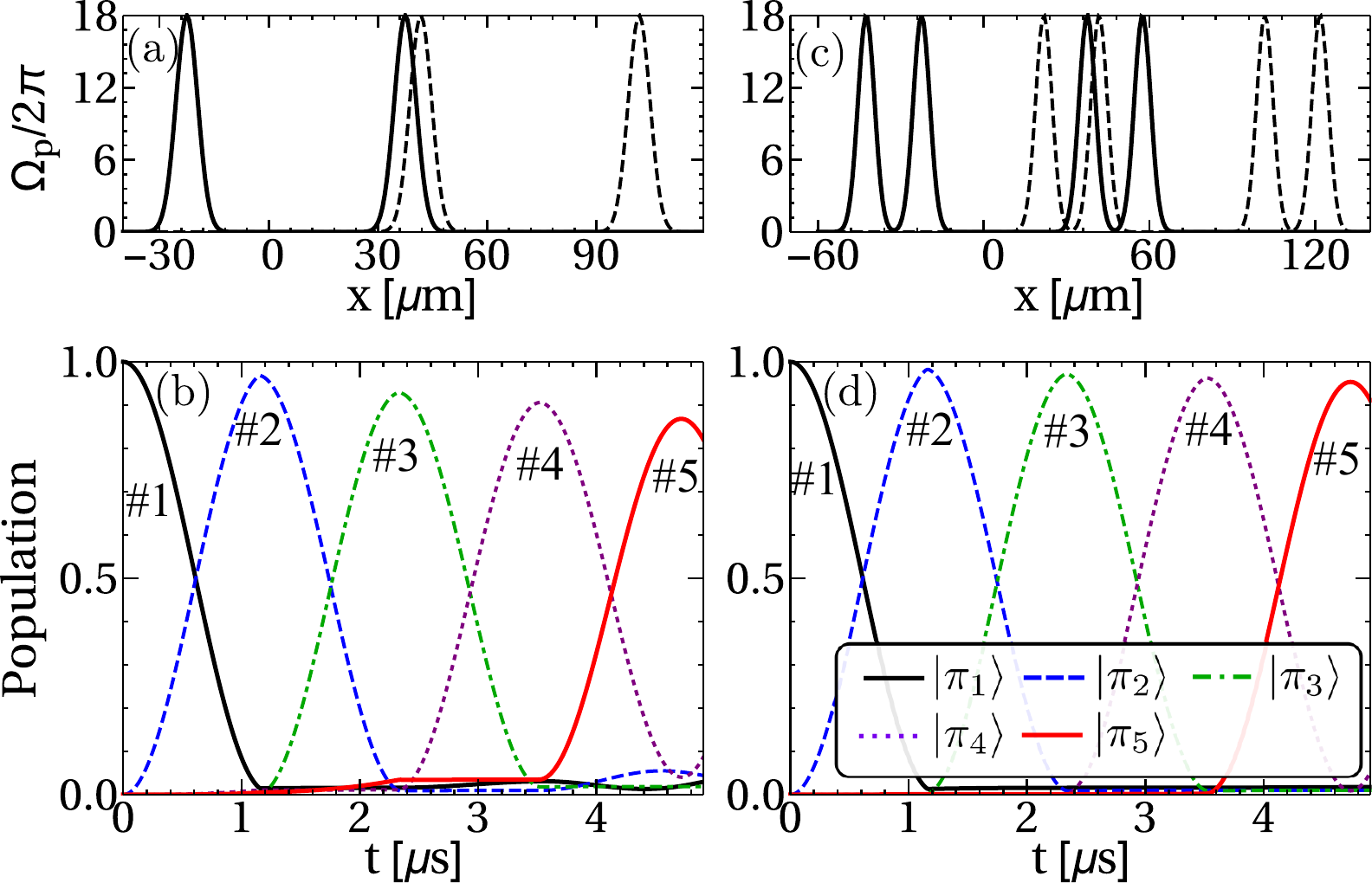}
	\caption{(a) The probe field $\Omega_p(x,t)$ with two Gaussian pulses as defined in \eref{omegap} for $(n=2)$, with peak Rabi frequency $\Omega_p$ corresponding to EIT-$\Xi$. Black solid (dashed) line indicates the initial (final) position of the polariton, induced by the spatially dependent coupling field  $\tilde{\Omega}_{c'}(x)$ shown in \fref{transport}~(b).  (b) Excitation $\rho_{nn}$ on each atom, with index $n$ indicated near each line, for $N=5$ Rydberg atoms shown in \fref{transport}~(a). The excitation is initialised on atom 1 with $\hat{\rho}=\ket{\pi_1}\bra{\pi_1}$ (black solid) and transported to atom 5 ($\ket{\pi_5}$, red solid) via atom 2 (blue dashed), 3 (green dot dashed) and 4 (purple dotted) following \eref{LB_eff}. (c-d) are similar to (a-b) but with four polaritons (\eref{omegap} with $n=4$).  The parameters used are the same as in \fref{transport}}
	\label{transport-2-4}
\end{figure}
Here we compare the fidelity in two cases: first with two polaritons initialized at position $x_1=-21.5$ $\mu$m and $x_2=38.5$ $\mu$m as discussed in  \fref{transport}~(a) of the main text, while in the second case we consider a set of four polaritons initialized at $x_1=-41.5$ $\mu$m, $x_2=-21.5$ $\mu$m, $x_3=38.5$ $\mu$m and $x_4=58.5$ $\mu$m, as presented in \fref{transport-2-4}~(c). In either case, we initialize the aggregate at site 1 $\hat{\rho}=\ket{\pi_1}\bra{\pi_1}$, and the system is then evolved with 
the effective master equation \eref{LB_eff}. We see that in either case, synchronization of the MIG by detector atoms with exciton transport can guide the exciton with high fidelity, but since four polaritons can suppresses long-range dipole-dipole interactions beyond the nearest neighbor, they yield a higher fidelity of up to 95\%.
 

\bibliographystyle{sebastian_v3}

\end{document}